\documentclass[amsmath, amssymb,floatfix,twocolumn,superscriptaddress,showpacs,showkeys,reprint,prb]{revtex4-1}
\usepackage{graphicx}
\usepackage{color}
\usepackage{verbatim}
\usepackage{textcomp}
\usepackage{mathbbol}
\usepackage[utf8]{inputenc}
\usepackage{float}
\usepackage{hyperref}
\usepackage{circuitikz}
\usepackage{tikz}
\usepackage{pgfplots}
\usepackage{amsmath,mathtools}

\pgfplotsset{compat=newest}
\usepgfplotslibrary{external} 
\usetikzlibrary{plotmarks}

\newlength\fheight 
    \newlength\fwidth

\newcommand{\diagentry}[1]{\mathmakebox[1.8em]{#1}}
\newcommand{\xddots}{
  \raise 4pt \hbox {.}
  \mkern 6mu
  \raise 1pt \hbox {.}
  \mkern 6mu
  \raise -2pt \hbox {.}
}

\begin{document}
    \setlength\fheight{0.55\columnwidth} 
   \setlength\fwidth{0.7\columnwidth} 

\title{Depinning of disordered bosonic chains}

\author{Nicolas \surname{Vogt}}
\affiliation{Institut f\"ur Theorie der Kondensierten Materie,  \\ Karlsruhe Institute of Technology, D-76128 Karlsruhe, Germany}
\affiliation{Chemical and Quantum Physics, School of Applied Sciences, RMIT University, Melbourne, 3001, Australia}
\author{Jared H. \surname{Cole}}
\affiliation{Chemical and Quantum Physics, School of Applied Sciences, RMIT University, Melbourne, 3001, Australia}
\author{Alexander \surname{Shnirman}}
\affiliation{Institut f\"ur Theorie der Kondensierten Materie,  \\ Karlsruhe Institute of Technology, D-76128 Karlsruhe, Germany}
\affiliation{L. D. Landau Institute for Theoretical Physics RAS, 
Kosygina street 2, 119334 Moscow, Russia}

\begin{abstract}
We consider one-dimensional bosonic chains with a repulsive boson-boson interaction that decays exponentially on large length-scales. 
This model describes transport of Cooper-pairs in a Josepshon junction array, or transport of magnetic flux quanta in quantum-phase-slip ladders, i.e. arrays of superconducting wires in a ladder-configuration that allow for the coherent tunnelling of flux quanta. 
In the low-frequency, long wave-length regime these chains can be mapped to an effective model of a one-dimensional elastic field in a disordered potential. 
The onset of transport in these systems, when biased by external voltage, is described by the standard depinning theory of elastic media in disordered pinning potentials. 
We numerically study the regimes that are of relevance for quantum-phase-slip ladders. 
These are (i) very short chains and (ii) the regime of weak disorder. 
For chains shorter than the typical pinning length, i.e., the Larkin length, the chains reach a saturation regime where the depinning voltage does not depend on the decay length of the repulsive interaction. 
In the regime of weak disorder we find an emergent correlation length-scale that depends on the disorder strength.
For arrays shorter than this length the onset of transport is similar to the clean arrays, i.e., is due to the penetration of solitons into the array.
We discuss the depinning scenarios for longer arrays in this regime.

\end{abstract}

\date{\today}
\maketitle

\section{Introduction}

Depinning theory describes the onset of propagation in many different physical systems. Examples include electrical transport in charge density waves\cite{Fukuyama1978,Brazovskii2004}, the critical current of magnetic flux lattices\cite{Larkin1973,Larkin1979} in type II superconductors, the propagation of magnetic domain boundaries\cite{Imry1975} and  crack formation in strained materials\cite{Laurson2010}. It was recently shown that the onset of electrical transport in one-dimensional arrays of Josephson junctions is also determined by the depinning of the charge-configuration along the array\cite{Vogt2014b}.

In this paper we consider a more general model, a discrete chain occupied by Bosons with a repulsive interactions that decays exponentially on long length-scales. In such a model the interaction between neighbouring islands can be expressed by introducing a continuous variable, quasi-charge/flux, whose value is determined by the distribution of bosons along the chain. Assuming that the continuous variable changes adiabatically, an effective model can be derived with the help of the Born-Oppenheimer approximation \cite{Homfeld2011,Vogt2014b}. In the case that disorder is present in the system, depinning theory can be applied to find the critical driving force that leads to a steady boson transport through the system. The Josephson junction arrays studied in Ref.~\onlinecite{Vogt2014b} represent a particular realization of this 
model. Alternatively, our results describe the dual system of quantum phase slip ladders. 
In the latter case (QPS ladders) the bosons are magnetic flux quanta that tunnel through quantum phase slip elements\cite{Mooij2005,Mooij2006,Mooij2014} that separate the loops in a ladder.

In voltage biased Josephson junction arrays, the depinning theory describes the transition from an insulating regime at low voltages to a transport regime at higher voltages. The critical voltage of the transition is referred to as the switching voltage. The insulating regime of the arrays is governed by an effective sine-Gordon-like quasi-charge model. 

In the study of the onset of transport in Josephson junction arrays, Ref.~\onlinecite{Vogt2014b}, the connection to standard depinning theory was established under the assumption of strongly disordered background charges (also referred to as the maximal disorder model). Under this assumption the disorder-term in the  effective model is spatially uncorrelated, allowing one to apply the standard depinning theory. Additionally the mapping to the standard depinning theory assumes the array length $N$ to be much larger than all other length-scales of the problem.

In this paper we study the depinning behavior of more general chain models in a parameter regime that is not found in Josephson junction arrays. Specifically we consider chains which do not meet one of the two aforementioned conditions, e.g. the chains are either short or only weakly disordered. In the case of short chains we find a saturation regime. In the case of weak disorder, we find spatially correlated long-range disorder in the effective sine-Gordon-like model. In the considered chain model the spatial correlation in the disorder decays approximately exponentially for large distances. The depinning process in systems with spatial correlations that decay with a power law has been studied in Ref.~\cite{Fedorenko2006} with the help of functional renormalization group theory. 

\section{Theory}

\subsection{The model}

We consider a chain of islands,
\begin{align}
	H &= \sum_{i,j} \frac{1}{2} \tilde n_i M_{i,j} \tilde{n}_j - \sum_{i} t \left( b_{i+1}^{\dagger}b_i^{\phantom \dagger} + \textrm{h.c.}\right)   \\
  \tilde n_i &= N_i - N_0 - f_i\ ,
\end{align}
where $N_i$ are the discrete bosonic occupation numbers of the islands, $N_0$ is the average occupation number at equilibrium (chemical potential, positive charge background), $f_i$ are the random gate charges, $b_i$, $b_i^{\dagger}$ are the bosonic annihilation and creation operators ($N_i = b_i^{\dagger}b_i$) and the bosonic tunnelling amplitude is given by $t$. 
In the limit $N_0 \gg 1$ we can replace $b_i \approx \sqrt{N_0} e^{-i\varphi_i}$, $b^{\dag}_i \approx \sqrt{N_0} e^{i\varphi_i}$, which leads to 
\begin{align}
	H &= \sum_{i,j} \frac{1}{2} \tilde n_i M_{i,j} \tilde{n}_j - \sum_{i} E_t \cos(\varphi_{i-1}-\varphi_i)\ ,  
\end{align}
where $E_t\equiv N_0 t/2$. For convenience we also introduce $n_i \equiv N_i - N_0$, so that $\tilde n_i = n_i - f_i$.

We assume that the long-range behavior of the interaction matrix $M_{i,j} = M_{|i-j|}$ is determined by an exponential decay on a length-scale $\Lambda$,
\begin{align}
	M_{i,j} &\propto e^{-\frac{|i-j|}{\Lambda}} \qquad \textrm{for} \ \ |i-j| \ge \Lambda
\end{align}

One concrete example of this situation is a Josephson array, a chain of superconducting island coupled via Josephson junctions and with self-capacitances $C_0$ (capacitances to the ground) and capacitances $C$ between the neighboring islands (junction capacitances). 
In this case $E_t = E_J$ is the Josephson energy whereas the coupling matrix is given by
\begin{align}
M_{i,j} =(2e)^2 \left[\hat C^{-1}\right]_{i,j}\ , 
\end{align}
where 
\begin{align}\label{CMatr}
\hat C = \begin{pmatrix}
 &-C\\
-C &C_0+2C & -C\\
&-C &C_0+2C & -C\\
&\qquad \diagentry{\xddots} &\qquad \diagentry{\xddots} &\qquad \diagentry{\xddots}\\
\end{pmatrix} \ .
\end{align}
This gives for the Fourier transform $M(k) \equiv \mathcal{FT}(M_{i-j})$ and for $M_{|i-j|}$ the following expressions:
\begin{align}
        &M(k) =  \Lambda E_C \frac{\frac{2}{\Lambda}}{\frac{1}{\Lambda^2}+2\left(1-\cos(k)\right)}\ , \\
	&M_{|i-j|} \approx \Lambda E_C e^{-\frac{|i-j|}{\Lambda}} \ ,
\end{align}
where the junction charging energy is defined by $E_C\equiv(2e)^2/2C$. In particular the energy cost of a single charge 
in such an array is of the order $M_{j,j} \approx \Lambda E_C$. Activated behavior with activation energy of 
order $\Lambda E_C$ was observed in Ref.~\onlinecite{Zimmer2013}.

It has been realized long ago~\cite{Korshunov1989,Choi1998} 
that for $\Lambda \gg 1$ the junction 
variables provide a more appropriate description than the island ones. We introduce, thus
\begin{align}
  m_i = \sum_{j=1}^{i-1} n_j  \quad,\quad
  F_i =  \sum_{j=1}^{i-1} f_j  \label{eqn:jja_qps:def_mF} \ .
\end{align}
The resulting Hamiltonian has the form,
\begin{align} \label{eq:Hm}
	H =\frac{1}{2}  \sum_{i,j} \left(m_i - F_i \right) U_{i,j} \left(m_j - F_j \right) 
	- \sum_{i} E_t \cos(\theta_i)\ ,
\end{align}
with the modified coupling matrix 
\begin{align}
U_{i,j} = U_{|i-j|} = 2 M_{|i-j|}-M_{|i-j|-1}-M_{|i-j|+1} \ .
\end{align}
Here $\theta_i \equiv \varphi_{i-1}-\varphi_i$.
One can easily check that $m_i$ and $\theta_i$ are conjugate variables.

A qualitative picture in the low energy, long wave length regime can be obtained from the Fourier transform 
of the coupling matrice,
\begin{align}
	U(k) &= \mathcal{FT}(U_{i-j}) =  2 (1-\cos(k))M(k) \ ,
\end{align}
where $k \in \left[-\pi,\pi\right]$.
The $k\rightarrow 0$ behavior of $M_{|i-j|}$ is dominated by the exponential decay and for small $k \ll \Lambda^{-1}$ the Fourier transform of the interaction matrix $M_{|i-j|}$ is approximately constant, $M(k) \approx M(0)$, which leads to
\begin{align}
	U(k) \approx M(0) k^2 \ .
\end{align}

\subsection{Standard Villain transformation}

The model (\ref{eq:Hm}) can be treated by the standard technique involving Villain approximation~\cite{Choi1998}.
We omit all the details and only mention the fact that the spin wave part of the resulting model in the limit $k\ll \Lambda^{-1}$, 
where $U(k) \approx M(0) k^2$ is quadratic, corresponds to a Luttinger liquid \cite{Giamarchi2004},
\begin{align}
	&H =\frac{v}{2} \int \textrm{d}x \left[K \left[\theta(x)\right]^2 + \frac{1}{K} \left[\partial_x m(x)\right]^2\right]\ , \\
	&\left[\theta(x_1),m(x_2)\right] = i \delta(x_1-x_2) 	\, , \label{eqn:LuttingerLagrangian} 
\end{align}
where the Luttinger liquid velocity $v$ and the Luttinger liquid parameter $K$ scale with the original model parameters as
\begin{align}\label{LuttingerParameters}
	v &= \sqrt{M(0) E_t}\ , \\
	K &= \sqrt{\frac{ E_t}{M(0)}} \, .
\end{align}
The corrections to (\ref{eqn:LuttingerLagrangian}) due to vortices (phase slips) in this type of 
theories (see, e.g., Ref.~\onlinecite{PhysRevB.16.1217})
are of the form $\propto \cos(2\pi p [m(x)+F(x)])$ with $p \in \mathbb{Z}$. The amplitude in front of this terms 
(fugacity of vortices) is predicted to be small in the limit $\Lambda \gg 1$~\cite{Korshunov1989,Choi1998} so 
that without disorder the critical value of $K$ is close to $2/\pi$ (it may be renormalized by disorder~\cite{Giamarchi2004}). 

We assume that $M(0) \gg E_t$ and therefore $K \ll 2/\pi$.
In this case the system is firmly in the charge density wave (CDW)-regime\cite{Giamarchi2004} and dominated 
by classical charge dynamics. The disorder $F(x)$ enters the relevant phase slip terms $\propto \cos(2\pi [m(x)+F(x)])$ and 
can pin the charge density profile.

\subsection{Alternative derivation}

Here we generalize the derivation given in Ref.~\onlinecite{Homfeld2011} for the case of a Josephson array described 
by the capacitance matrix (\ref{CMatr}) to the case of a more general matrix $M_{i,j}$ characterized by a screening 
length $\Lambda$. 
We start by rewriting the Hamiltonian (\ref{eq:Hm}) as
\begin{align}
	H =&\frac{1}{2}  \sum_{j} U_0 \left(m_j - F_j \right)^2  - \sum_{j} E_t \cos(\theta_j) \ , \nonumber\\
	&+\frac{1}{2}  \sum_{i,j} \left(m_i - F_i \right) \delta U_{i,j} \left(m_j - F_j \right)\ ,
\end{align}
where $U_0 = U_{j,j}$ and $\delta U_{i,j} = U_{i,j} - U_0 \delta_{i,j}$. 
Next we transform  the third term with the help of the Hubbard-Stratonovich transformation, 
which introduces a new degree of freedom $Q_i$, often referred to as the quasi-charge. This gives 
\begin{align}
	H\{Q\} =&\frac{1}{2}  \sum_{j} U_0 \left(m_j - F_j \right)^2 - \sum_{j} E_t \cos(\theta_j) \ , \nonumber\\
	&-U_0\sum_{j} Q_j (m_j-F_j) - \frac{U_0^2}{2}  \sum_{i,j}  Q_i \left[\delta U^{-1} \right]_{ij}Q_j\ , 
\end{align}
such that $H = min_Q[H\{Q\}]$. Next
\begin{align}
	H\{Q\} =&\frac{1}{2}  \sum_{j} U_0 \left(m_j - F_j -Q_j\right)^2 - \sum_{j} E_t \cos(\theta_j) \nonumber\\
	& + \frac{1}{2}  \sum_{i,j}  Q_i B_{i,j} Q_j\ ,
\end{align}
where $B_{i,j}=-U_0^2\left[\delta U^{-1} \right]_{ij} - U_0\delta_{i,j}$.
The Fourier transform reads
\begin{align}
B(k) = -U_0 - \frac{U_0^2}{U(k)-U_0}\ .
\end{align}
For small wave vectors $k\ll \Lambda^{-1}$ we obtain $U(k)\ll U_0$ and $B(k) \approx U(k)\approx M(0) k^2$.
Thus, assuming $Q_i$ changes slowly enough as a function of the coordinate $i$, i.e., changes on length scales longer than $\Lambda$, we can approximate 
\begin{align}\label{HQ_approx}
	H\{Q\} \approx &\frac{1}{2}  \sum_{j} U_0 \left(m_j - F_j -Q_j\right)^2 - \sum_{j} E_t \cos(\theta_j)  \nonumber\\
	& + \frac{M(0)}{2}  \sum_{i}  (Q_i - Q_{i+1})^2\ .
\end{align}

For the Josephson arrays with the capacitance matrix (\ref{CMatr}) we obtain
$U_0 = 2 E_C$ and $M(0) = 2\Lambda^2 E_C = 2 E_{C0}$, where 
$E_{C0}\equiv (2e^2)/2C= \Lambda^2 E_C$. 
In this case the form of the third term of (\ref{HQ_approx}) is exact~\cite{Homfeld2011}. 

The adiabatic dynamics of the model (\ref{HQ_approx}) without disorder was 
analyzed in Ref.~\onlinecite{Homfeld2011}. The inclusion of disorder is straightforward. 
The aim is to integrate out the degrees of freedom $(m_i,\theta_i)$.
For a given (adiabatic) trajectory $Q_i(t)$ the dynamics factorizes to independent dynamics 
of single junctions governed by the Hamiltonians 
\begin{equation}\label{SingleJJHam}
H_i(Q_i) = \frac{1}{2}\,U_0 \left(m_i - F_i - Q_i\right)^2 - E_t \cos(\theta_i) \ . 
\end{equation}
The Born-Oppenheimer periodic potential is given by the ground state of the well known 
Hamiltonian (\ref{SingleJJHam}), $E_Q(Q_i + F_i)$, where $Q_i+F_i$ serves here as the 
total quasi-charge. The function $E_Q(Q)$ is periodic with period $1$ as can be seen from 
(\ref{SingleJJHam}). In the limit $E_t \gg U_0$ ($E_J \gg E_C$) it is given by 
$E_Q(Q) = E_S \cos(Q)$, where $E_S$ is the quantum phase slip amplitude~\cite{PhysRevLett.89.096802}. 

Thus we obtain the effective potential energy of the whole array of the form 
\begin{align}
	U_C &=\frac{1}{2}  \sum_{i,j}  Q_i B_{i,j} Q_j  + \sum_j E_Q(Q_i + F_i) \nonumber\\
	&\approx \frac{1}{2}\sum_i M(0)(Q_i - Q_{i+1})^2 + \sum_i E_Q(Q_i + F_i) \nonumber\\
	&=\sum_i E_{C0}(Q_i - Q_{i+1})^2 + \sum_i E_Q(Q_i + F_i)\ .
\end{align}
This potential is supplemented by the kinetic energy. In the limit $E_t \gg U_0$ ($E_J \gg E_C$)  it reads 
$T = (1/2)\sum_i L \dot Q_i^2$, where the $L$ is the Josephson inductance $L \approx L_J =1/ E_t=1/E_J$.
The quadratic part of the Lagrangian $T - U_C$ gives again the Luttinger liquid with the parameters 
(\ref{LuttingerParameters}). Since we assume $K\ll 2/\pi$, the periodic potential $E_Q(Q_i + F_i)$ is relevant and 
pins the density profile. In what follows we investigate the charge pinning in this setup.

\subsection{Edge bias}

From now on we employ the terminology of Josephson junction arrays and put $2e =1$.
As we are primarily interested in the transport properties of the chains, we introduce a bias $V$ at the edge:
\begin{align}
	U_C &= \sum_{i} \frac{\left(Q_i-Q_{i+1}\right)^2}{2C_0} + E_Q(Q_i+F_i)  -\frac{V}{C} Q_1 \ .
\end{align}
To simplify the treatment in terms of the depinning theory we transform the system from a boundary biased situation to a spatially homogeneous driving by introducing a parabolic shift in $Q$ and $F$,
\begin{align}
	\tilde{Q}_i &= Q_i - \frac{C_{0}}{C} V  \frac{(N+1-i)(N-i)}{2N} \ , \\
	\tilde F_i &= F_i + \frac{C_{0}}{C} V  \frac{(N+1-i)(N-i)}{2N} \ ,
\end{align}
and the corresponding potential part of the Hamiltonian with a driving force $V/N$,
\begin{align}
	U_C &= \sum_{i}  \frac{\left(\tilde Q_i- \tilde Q_{i+1}\right)^2}{2 C_0} + E_Q\left( \tilde Q_i + \ \tilde F_i\right) + \frac{V}{N} \frac{\tilde Q_i}{C} \label{eqn:Q_Ham} \ .
\end{align}

In this formulation the problem corresponds to the discrete version of the  well known depinning problem in one-dimension\cite{Brazovskii2004}. The elastic energy of the field $Q_i$ is determined by the elastic constant $C_{0}$. The elastic field is pinned by the random pinning potential $E_Q(\tilde Q_i + \ \tilde F_i)$ and driven by the homogeneous driving force $V/N$. In the pinned regime the applied force $V/N$ is not strong enough to overcome the potential barrier imposed on the elastic object $Q_i$ by the random pinning potential. 

In the case that no driving force is applied, $V=0$, the form of the elastic object is determined by a competition between the elastic term $(Q_i-Q_{i+1})^2$ and the pinning term $E_Q(Q_i+F_i)$ in $U_C$. On small length-scales, where the elastic energy term dominates, $Q_i$ is approximately constant. The field $Q_i$ changes on large length-scales where the pinning potential dominates. The crossover between the two regimes happens at the length scale $L_p$, which was first determined by Larkin for a flux line lattice in type II superconductors \cite{Larkin1979}. The length $L_p$ goes by many names depending on the physical systems that are pinned. In type II superconductors it is called Larkin length, in ferromagnets with domain boundaries Imry-Ma length \cite{Imry1975} and for charge density waves it is called Fukuyama-Lee length \cite{Fukuyama1978}. In this work we use the term Larkin length.

Once the driving force $V/N$ exceeds a critical force $V_{cr}/N$, the pinning potential is overcome and the elastic object starts to move through the disordered medium. An intuitive argument to find the value of the critical driving force can be found by comparing the driving force to the pinning force at the Larkin length\cite{Larkin1979}. The distribution of $Q$ is rigid on length-scales up to the Larkin length. The elastic object can only start to move when the driving force exceeds the collective pinning force on a segment with length $L=L_{p}$.

\subsection{Strong disorder}
We first consider the strongly disordered model for which the results of the standard depinning theory\cite{Fukuyama1978,Brazovskii2004} are directly applicable. To make the connection to these results the difference between the original disorder ($f_i$) and the effective quasi-disorder before ($F_i$) and after ($\tilde F_i$) the parabolic shift in the quasicharge is important. In terms of the original disorder the strongly disordered model is defined by,
\begin{align}
	f_i &\in \left[-1/2,1/2\right] \ , \\
	p(f_i) &= \Theta_H\left(\frac{1}{2} - \left| f_i\right|\right) \ ,
\end{align}
where $p(f_i)$ is the probability distribution of the disorder $f_i$ and $\Theta_H$ is the Heaviside $\Theta$-function. This model corresponds to the strongest possible disorder in the considered chain model. A frustration $f_i$ with an absolute value larger than $1/2$ is compensated by placing an additional (anti)-boson on the $i$-th island of the chain. The disorder is bounded by $\pm 1/2$ and a box-distribution of disorder-charges inside these boundaries is the maximal disorder. 
While $f_i$ itself is not spatially correlated, in the effective quasi-charge model, the quasi-disorder $\tilde F_i$ is correlated between different islands $i$ and $j$,
\begin{align}
	\left\langle \tilde F_i \tilde F_j\right\rangle_{dis} \ne 0 \quad \textrm{for} \quad i \ne j  \ .
\end{align}
At first this seems to deviate from the normal situation in depinning theory where the disorder in the system is not spatially correlated\cite{Brazovskii2004}. However,  in the depinning theory, only correlations in the pinning potential are relevant to the behaviour of the system.
The  potential $E_Q$ is a function of the quasi-charge with a periodicity of $1$. Since the disorder $f_{i}$ is box distributed in an interval that corresponds to the periodicity of the potential, the offset $\tilde F_{i}$ can be absorbed into another uncorrelated box-distributed disorder term $f^b_i$,
\begin{align}
	&\tilde F_{i} =  F_{i-1} + \frac{C_{0}}{C} V  \frac{(N+1-i)(N-i)}{2N} +f_{i} \rightarrow f^b_i \ , \\ 
	&E_{Q}\left(\tilde Q_i +   \tilde F_{i}  \right) \rightarrow E_{Q}\left(\tilde Q_i + f^b_{i} \right) \ , \\
	&f^b_i \in \left[-\frac{1}{2},\frac{1}{2}\right] \ , \\
	&p(f^b_i) = \Theta_H\left(\frac{1}{2} - \left| \tilde f_i\right|\right)  \ .
\end{align}
From the point of view of the potential $E_Q$, the quasi-disorder $\tilde F_i$ is equivalent to a spatially uncorrelated disorder term $f^b_i$ in the maximally disordered model. 

Another way to determine whether spatial correlations in $\tilde F_i$ affect the quasi-charge model, is to calculate the disorder-averaged correlation function of the pinning potential: 
\begin{align}
	\left\langle E_{Q}\left(\tilde Q_1 + \tilde F_i\right)  E_{Q}\left(\tilde Q_2 + \tilde F_j\right)\right\rangle_{dis} &=  R(Q_2-Q_1) \delta_{i,j} \ ,
\end{align}
where the correlation function $R(Q)$ is given by,
\begin{align}
	R(Q) =  \int_{-\frac{1}{2}}^{\frac{1}{2}} \textrm{dF} \ E_{Q}(Q+ F ) E_{Q}(F) \ .
\end{align}
Since the correlator of the pinning potential is proportional to a Kronecker delta, the pinning potential is not spatially correlated.

We have now seen that for the maximal disorder model we arrive at an effective model that conforms with the standard assumptions of depinning theory. In this case the Larkin length and the critical driving force are well known (see for example Ref. \onlinecite{Fukuyama1978}). 

The approximate value of the Larkin length in one-dimensional systems is given by \cite{Fukuyama1978},
\begin{align}
	L_{p} &= 3^{-\frac{2}{3}} \Lambda^{\frac{4}{3}} \left[\tilde{R}\left(\frac{E_J}{E_C} \right)\right]^{-\frac{2}{3}}\ .
\end{align}
The relevant parameters of the chain are the  energy $E_C$, the tunnelling amplitude $E_J$, the chain length $N$ and $\Lambda$. To express the Larkin length in terms of these parameters we have defined the function $\tilde R$,
\begin{align}
	\tilde{R}\left( \frac{E_J}{E_C} \right) &= \left(\frac{\left(E_{Q}^{\textrm{max}}\right)^2}{16 E_C^2}\right) \ ,
\end{align}
where $E^{\textrm{max}}_Q$ is the amplitude  of the random pinning potential $E_Q(\tilde Q_i+\tilde F_i)$.
The correlation function $\tilde R$ is a function of the dimensionless ratio of the tunnelling matrix element and $E_C$. The function needs to be determined numerically only once for all possible values of chain length and $C_{0}$.

Similarly the depinning force can be expressed in terms of $\tilde R$ and is given by\cite{Brazovskii2004},
\begin{align}
	V_{cr} &\approx N \frac{1}{C_{0}} l \frac{1}{L_p^2} \label{eqn:depinning:depinning_force} \\
	&= N  
	3^{\frac{4}{3}} \, \Lambda^{-\frac{2}{3} }  \left\{ \tilde{R}\left( \frac{E_J}{E_C}\right)\right\}^{\frac{2}{3}} \ . \label{eqn:depinning:sw_Voltage} 
\end{align}
Further corrections to this intuitive approach can be obtained from renormalization-group-approaches\cite{Brazovskii2004, Chauve2000a, Chauve2001}. We use the approximation Eq.\ref{eqn:depinning:depinning_force} in this work.

In most systems, where depinning theory is applicable, the system size is much larger than the Larkin length and it is a good approximation to assume infinite system size. We now turn our attention specifically to short finite chains. From 
Eq.~\ref{eqn:depinning:sw_Voltage} we see that the critical driving force decreases with increasing $\Lambda$. At the same time the Larkin length increases,
\begin{align}
	L_p &\propto \Lambda^{\frac{4}{3}} \ .
\end{align}
In finite chains the Larkin length becomes equal to the system size $N$ when $\Lambda$ reaches the value,
\begin{align}
	\Lambda_N &= N^{\frac{3}{4}} 3 \left\{\tilde R\left(\frac{E_J}{E_C}\right)\right\}^{\frac{1}{4}} \ .
\end{align}
Increasing $\Lambda$ further while keeping $E_C$ constant only increases the elastic energy $E_{C0}$ coupling neighbouring islands. The Larkin length, the length-scale on which $Q$ is approximately constant, should increase. However in this limit the Larkin length is equal to the system size and therefore the field $Q$ is approximately constant along the whole array.  

For $\Lambda \gg \Lambda_N$ the critical driving force is independent of the interaction length as long as $E_C$ is kept constant. A lower boundary for $V_{cr}$ is approximately given by
\begin{align}
	V_{cr} &\approx \sqrt{N} 
	3^{-\frac{1}{2}} \left\{ \tilde{R}\left( \frac{E_J}{E_C}\right)\right\}^{\frac{1}{2}} \ , \label{eqn:depinning:sw_Voltag_short_array} 
\end{align}
which is the critical driving  $V_{cr}$ one finds for  $\Lambda = \Lambda_N$. In reality $V_{cr}$ saturates for smaller $\Lambda$, when $N$ is of the same order of magnitude as $L_{p}$ (for comparison see the numerical simulations in 
Sec.~\ref{sec:depinning:numeric}). This leaves the principal behaviour of Eq.~\ref{eqn:depinning:sw_Voltag_short_array} unchanged and contributes a prefactor of order one in the expression for the critical driving force.  

\subsection{Weak disorder}
In the weak disorder case the bare disorder $f_i$ is not evenly distributed in the interval $\left[-1/2,1/2\right]$.
One particular system we have in mind is a ladder of quantum phase slip (QPS) junctions~\cite{Mooij2006,Mooij2014}. In such a system, superconducting wires are arranged in a ladder configuration, such that a one-dimensional chain of superconducting loops is formed. The superconducting wires that are shared by neighbouring loops contain a very thin section that forms the QPS-junction. Magnetic flux quanta in the loops assume the role of the bosonic particles. The QPS-junctions between the superconducting loops provide the hopping matrix element and the coupling matrix $M_{i,j}$ is the inverse inductance matrix of the system. In a ladder configuration of superconducting wires the inductance matrix has the exactly the previously mentioned tridiagonal form Eq.~\ref{CMatr}, as long as the kinetic inductance dominates over the geometric inductance.

Due to the lack of large magnetic dipoles in the vicinity of such a system, a weak disorder limit is more likely to be realized than in Josephson junction arrays.

We consider two models of weak disorder: (i) the weak box disorder
\begin{align}
  \ f_i &\in \left[-\frac{\gamma}{2} , \frac{\gamma}{2} \right] \ , \\
	p(f_i) &= \frac{1}{\gamma} \Theta_H\left(\frac{\gamma}{2} - \left| f_i\right|\right) \ ,
\end{align}
with the disorder strength $\gamma < 1$; (ii) Gaussian disorder
\begin{align}
p(f_i) &= \frac{1}{\sigma \sqrt{2 \pi}} e^{-\frac{1}{2} \frac{f_i^2}{\sigma^2}} \ ,
\end{align}
with a standard deviation $\sigma < 1/2$.

In the weak disorder models the spatial correlation in the quasi-disorder $\tilde F_{i+1}$ can not be neglected in the argument of the effective potential $E_Q(Q)$.  
The maximal value of the disorder  $ f_i$ is smaller than the periodicity of the potential $E_Q$ and the long range correlation in the quasi-disorder $\tilde F_i$ can not be absorbed in the potential. The correlation function of the pinning potential therefore acquires a long range correlation component. We decompose the correlation function into short and long-range components,
\begin{align}
	\left\langle E_{Q}\left(Q +\tilde F_i\right)  E_{Q}\left( \tilde F_j\right)\right\rangle_{dis} &= R(Q) \delta_{i,j}+ R_2(Q,i,j) \ ,	
\end{align}
with the $\delta$-correlated component $R(Q)$ and the long range correlation function $R_2(Q,i,j)$. Due to the long range correlations the intuitive picture of the depinning-transition is not valid anymore. For a long range correlation function, 
\begin{align}
	R_2(Q,i,j) \propto \left| i-j \right|^{-a} \ ,
\end{align}
that decays with a power law, the problem has been approached with the functional renormalization group method (FRG) in 
Refs.~\onlinecite{Fedorenko2006,Fedorenko2008}. 

It has been shown\cite{Santucci2010, Laurson2010} that these long-range correlations lead to the emergence of a new length-scale in the pinned system, the typical correlation length $L_{corr}$. The roughness function $w(x)$ of a pinned system shows a different behaviour, namely a variation in the roughness exponent $\zeta_{\textrm{rough}}$, depending on whether the system is probed at length-scales smaller or larger than the correlation length\cite{Laurson2010}. We derive typical correlation lengths for the two weak disorder models under the assumption that $E_{Q}$ can be approximated as a cosine-potential,
\begin{align}
	E_J &\sim E_C \ , \\
	E_{Q} (Q) &\approx E_{Q}^{max} \left[1-\cos\left(2 \pi Q\right)\right] \ .
\end{align}

To calculate the correlation function of the pinning-potential of two different chain sites $j$ and $k$ we set, without loss of generality, $j < k$. The correlation function in the weak box-disorder model is given by an integral over the disorder,
\begin{widetext}
\begin{align}
  R_2(Q,j,k) &= \left(E_{Q}^{max}\right)^2 \int_{-\infty}^{\infty} \textrm{d}F_j \tilde p(F_j) \left(\frac{1}{\gamma}\right)^{k-j} \int_{-\frac{\gamma}{2}}^{\frac{\gamma}{2}} \textrm{d}f_{j} \dots \int_{-\frac{\gamma}{2}}^{\frac{\gamma}{2}} \textrm{d}f_{k-1}  \cos\left(Y_1\right) \cos\left(Y_2\right) \ , \\
	 Y_1 &= Q +  F_j + V  \frac{(N+1-j)(N-j)}{2N} \ , \quad
	 Y_2  = Q +  F_j + \ \sum_{l=j}^{k-1} f_l +  V  \frac{(N+1-k)(N-k)}{2N}\ ,
\end{align}
\end{widetext}
where $\tilde p(F_j)$ is the probability distribution of the quasi-disorder $F_j$. Expressing the cosine in terms of exponentials one finds that the absolute value of the correlation function $R_2$ is bounded by an envelope function $R_E$,
\begin{align}
	&\left| R_2(Q,j,k) \right| \le R_E(Q,k-j) \nonumber\\
	&= 2 \ \left(E_{Q}^{max}\right)^2  \left(\frac{\sin\left(\pi \gamma\right)}{\pi \gamma} \right)^{k-j} \ .
\end{align}
The long-range correlation function decays exponentially with the distance $k-j$ and the correlation of the pinning-potential decays on the length-scale,
\begin{align}
	L_{corr} = - \frac{1}{\ln\left(\frac{\sin\left(\pi \gamma\right)}{\pi \gamma}\right)} \label{eqn:depinning:Lcorr_box} \ .
\end{align}
As expected the correlation length goes to zero in the limit of the maximal disorder and diverges in the clean limit without disorder,
\begin{align}
	\gamma &\rightarrow 1  \qquad \Rightarrow \qquad L_{corr} \rightarrow 0 \ ,  \\
	\gamma &\rightarrow 0  \qquad \Rightarrow \qquad  L_{corr} \rightarrow \infty \ . 
\end{align}
For a Gaussian distribution of the bare disorder $f_i$, the correlation function is
bounded by the exponential function $R_G$,
\begin{align}
  \left| R_2(Q,j,k) \right| \le R_G(Q,k-j) = 2 \ \left(E_{Q}^{max}\right)^2 \left(e^{-2 \pi^2 \sigma^2} \right)^{k-j} \ .
\end{align}
The correlation length is determined by the standard deviation $\sigma$ of the bare disorder,
\begin{align}
	L_{corr} &= \frac{1}{2 \pi^2 \sigma^2} \ .
\end{align}
We can again test the limits of infinitely broad and non-disordered distributions,
\begin{align}
	\sigma &\rightarrow \infty   \quad \ \  \Rightarrow  \qquad L_{corr} \rightarrow 0  \ ,  \\
	\sigma &\rightarrow 0   \qquad \Rightarrow \qquad L_{corr} \rightarrow \infty  \ . 
\end{align}
In the broad limit the Gaussian disorder shows the same asymptotic behaviour as the box-disorder distribution when approaching the maximal disorder limit. The maximal disorder limit is consistent with a very broad bare Gaussian distribution. In the opposite limit the Gaussian distribution corresponds to a homogeneous shift in the definition of the quasi-charge and the correlation length diverges.

 The correlation length $L_{corr}$ marks the crossover between a disorder free and a strongly disordered array. On length-scales smaller than the correlation length the value of the disorder $F_i$ is approximately constant and constitutes a mere shift in the field $Q$. If the weakly disordered system is probed on these length-scales it behaves like a clean chain. On larger length-scales the value of the disorder changes significantly and the system behaves like a disordered chain. This transition is shown in the next section with the example of the dependence of the threshold voltage on the length of the chains.

\section{Simulations}
\label{sec:depinning:numeric}
We obtain the critical driving force $V_{cr}$ by numerically solving the equations of motion of the field $Q_i$ that can be obtained from the Hamiltonian (\ref{eqn:Q_Ham}),
\begin{align}
  &\mathcal{M} \ddot{Q}_{i} + \frac{2 Q_{i}-Q_{i-1}-Q_{i+1}}{C_{0}} + \alpha_R \dot Q_i  \nonumber\\ &+ V_{Q}\left(Q_i +  F_i\right) = 0  \label{eqn:depinning:numerics_equatio_of_motion} \ , \\
  &\mathcal{M} \ddot{Q}_{1} + \frac{Q_{1}-Q_{2}}{C_{0}} + \alpha_R \dot Q_2  + V_{Q}\left(Q_1 \right) = \frac{V}{C} \ , \\
  &\mathcal{M} \ddot{Q}_{N+1} + \frac{Q_{N+1}-Q_{N}}{C_{0}} + \alpha_R \dot Q_{N+1}  \nonumber\\ &+ V_{Q}\left(Q_{N+1}  + F_{N+1}\right) = 0 \ .
\end{align}
The function $V_Q$ is the pinning force given by,
\begin{align}
	V_{Q}(Q) &\equiv \partial_Q E_Q(Q) \ .
\end{align}
To guarantee numerical convergence we have introduced a mass $\mathcal{M}$ and a linear dissipative term with a dissipation constant $\alpha_R$. Similar numerically simulations of the switching voltage in arrays of normal tunnel contacts have been conducted in 
Ref.~\onlinecite{Middleton1993}.

The critical driving force $V_{cr}$ is determined by adiabatically applying the boundary force $V$ and determining whether a stable solution for the field $Q_i$ can be found.  
Although $V$ is increased slowly, the switch-on time of the driving force $V$ in the numerical simulation is finite. The phenomenological dissipative term has to be included to compensate the small transport velocity $\dot Q_i$ introduced by the switch-on of $V$. The introduction of a phenomenological term is also a standard tool in the derivation of the depinning force $V_{cr}$ in renormalization-group-treatments of pinned systems \cite{Brazovskii2004}.  

The mass $\mathcal{M}$ and the dissipation parameter $\alpha_R$ both affect the dynamical properties of the system, however they have no influence on the breakdown of the static state. In the example of a Josephson junction array, the mass $\mathcal{M}$ corresponds to an inductance and $\alpha_Q$ corresponds to an Ohmic resistance. In a quantum phase slip ladder $\mathcal{M}$ corresponds to a capacitance.
In all simulations we choose the tunnelling amplitude and the coupling energy to be equal, $E_J = E_C$, so that the potential $E_Q$ is close to a cosine potential. The length of the chain $N$ and the parameter $\Lambda$ are varied.
\subsection{The clean chain}
\label{sec:depinning:numeric:clean}

\tikzexternalenable
\begin{figure}[!t]
	\centering
\includegraphics{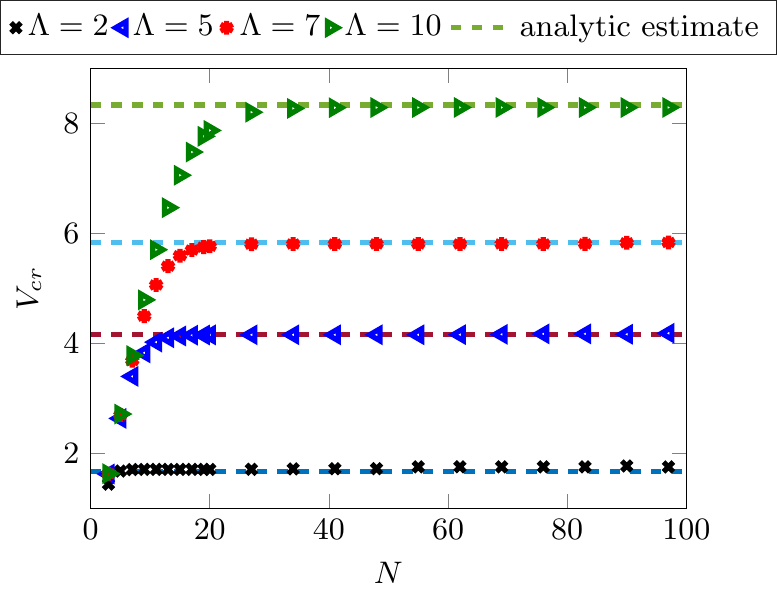}
\caption{\label{fig:depinning:Vt_N_nodisorder}(Color online) The critical driving force $V_{cr}$ of the clean chain is plotted as a function of the chain-length for several values of $\Lambda$. As long as the chain is more than twice as long as $\Lambda$, $V_{cr}$  is independent of the length $N$ and proportional to $\Lambda$. The critical driving force has the value predicted by an analytic estimate by Haviland and Delsing\cite{Haviland1996}.
In  the region where the chain is shorter than $\Lambda$ the system is in the zero-dimensional limit. The critical driving force is  proportional to $N$ and does not depend on $\Lambda$.  }
\end{figure}

\tikzexternalenable
\begin{figure}[!t]
	\centering
\includegraphics{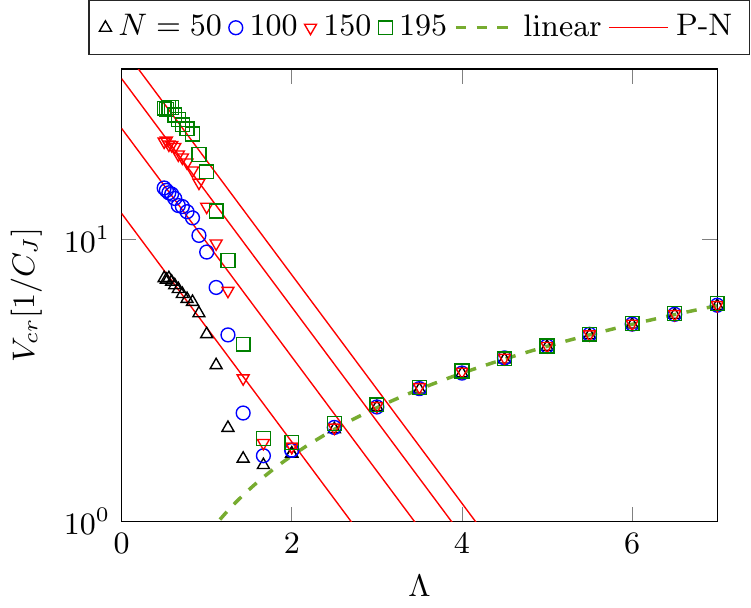}
\caption{\label{fig:depinning:Vt_lambda_nodisorder}(Color online) The critical driving force $V_{cr}$ as a function of  $\Lambda$ for different chain-lengths $N$ in the clean chain. As long as $\Lambda$ is larger than $2$, the analytic estimate Eq.~\ref{eqn:depinning:HDAnalyticEst} is reproduced and $V_{cr} \propto \Lambda$, as it is shown by the linear fit (dashed line) in the plot.
	For smaller $\Lambda$, non-propagating $Q$-excitations can be created in the chain by the adiabatic switch-on of the driving force. The number of  excitations is proportional to the chain-length and the critical driving is proportional to $N$-times the depinning-force of one excitation. The depinning force has been fitted (red lines) to an exponential function $V_{sol} = \beta e^{-\gamma \Lambda}$ as it arises from the Peierls-Nabarro-Potential\cite{Braun1998, Fedorov2011}. }
\end{figure}

We first simulate the clean model that has been used as  the default model in a number of experimental papers on Josephson junction arrays \cite{Haviland1996,Haviland2000,Agren2001}. While this model does not take into account charge disorder\cite{Johansson2000}, it might be more relevant for quantum-phase-slip-arrays than Josephson junction arrays as the former lack the strong charge disorder that can be found in the latter.   

In the clean case, a simple argument to determine the critical driving force $V_{cr}$ can be found in Ref.~\onlinecite{Haviland1996}.
In the continuum limit ($\Lambda \gg 1$) for long chains ($\Lambda \ll N$) the effective model of the clean chain is equal to the sine-Gordon model with a modified potential. The solutions of the standard sine-Gordon equation of motion are the well known solitons \cite{Homfeld2011,Fedorov2011},
\begin{align}
	Q(x) &= \frac{2}{\pi} \arctan\left(e^{\gamma_{sol} \frac{x-vt}{\Lambda}}\right) \ , \\ 
	\gamma_{sol} &= \frac{1}{\sqrt{1 - \frac{v^2}{L C_{0}}}} \ ,
\end{align}
with the soliton velocity $v$. The  spatial derivative of a static soliton $v=0$ has a maximal value of
\begin{align}
	\left. \partial_x Q(x) \right|_{v=0} \le \frac{1}{\pi} \frac{1}{\Lambda} \ .
\end{align}
The boundary driving force $V$ takes the form of a boundary condition on the spatial derivative at $x=0$,
\begin{align}
	\left. \partial_x Q(x) \right|_{x=0} &= \frac{C_{0}}{C} V \ . 
\end{align}
This can be used to estimate the maximal boundary force $V$ for which a static soliton can exist at the array ends,
\begin{align}
  V_{cr} &= \frac{4}{\sqrt{\pi}} \sqrt{\frac{C}{2 C_{0}} V_Q^{max}} \propto \Lambda  \label{eqn:depinning:HDAnalyticEst} \ , \\
	V_Q^{max} &= \max_{Q}\left(\partial_Q E_Q(Q)\right)\ .
\end{align}
In the Josephson junction arrays this force corresponds to the  switching voltage at which the array switches from insulating to transport behaviour. 

The critical driving force does not depend on the array length and is proportional to the interaction length $\Lambda$. Both features are confirmed by the numerical simulations in Fig.~\ref{fig:depinning:Vt_N_nodisorder} and Fig.~\ref{fig:depinning:Vt_lambda_nodisorder}.

In the limit $\Lambda > N $  the spatial dependent field $Q_i$ takes the same value on all islands of the chain, $Q_i \rightarrow Q$ and the coupled equations of motion simplify to a single equation of motion,
\begin{align}
  &\mathcal{M} \ddot{Q}  + \alpha_R \dot Q  + V_{Q}\left(Q \right) = \frac{V}{C} \ .
\end{align}
The one-dimensional clean chain model reduces to a zero-dimensional model. The critical driving force increases linearly with array size and is independent of $\Lambda$ (Fig.~\ref{fig:depinning:Vt_N_nodisorder}).

When the interaction length $\Lambda$ is comparable to the inter-site distance $\Lambda < 2 $ we are no longer in the continuum limit and the analytic approximation Eq.~\ref{eqn:depinning:HDAnalyticEst} is not valid. 
The switching-voltage is proportional to the length $N$ and the $\Lambda$-dependence can be fitted to an exponential behaviour,
\begin{align}
  \label{eqn:P_N}
	V_{cr} &= N \beta e^{-\gamma \Lambda} \ ,
\end{align}
as seen in Fig.~\ref{fig:depinning:Vt_lambda_nodisorder}. Only one set of fitting parameters $\beta, \gamma$ is used for all four simulated chain-lengths.

The change of the switching voltage behaviour can be understood in the following way. The interaction length $\Lambda$ is a measure for the ratio of the elastic coupling between neighbouring islands and the depth of the pinning-potential. For small interaction lengths $\Lambda < 2$,  $Q$-excitations can be created at the driven end of the chain without leading to complete depinning.
During the adiabatic increase of force $V$ the whole chain is filled with non-propagating  $Q$-excitations. The depinning transition of these $Q$-excitations is determined by the Peierls-Nabarro-potential\cite{Braun1998,Fedorov2011}.
This give rise to a $\Lambda$-dependence of the form of Eq.\ref{eqn:P_N}.
In the context of Josephson-junction-arrays this was discussed by Fedorov et al. for the depinning of a single $2e$-charge-excitation \cite{Fedorov2011}.
\subsection{The maximally disordered array}
\label{sec:depinning:numeric:disorder}

\tikzexternalenable
\begin{figure}[!t]
	\centering
\includegraphics{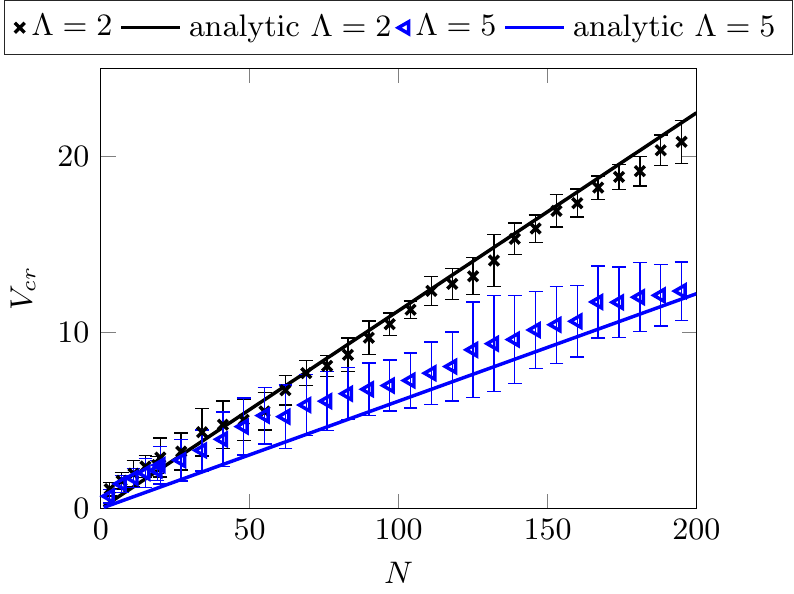}
\caption{\label{fig:depinning:Vt_N_disorder}(Color online) The critical driving force $V_{cr}$ is plotted as a function of the length $N$ of the disordered chain. For $\Lambda = 2$ (black crosses) the critical force grows linearly with the chain-length $N$ as expected from the analytic estimate (black line). For $\Lambda =5$ (blue triangles) $V_{cr}$ is proportional to $N$ at larger chain-lengths when $N \approx 100 \approx 2.5 L_{p}$ and fits the analytic estimate (blue solid line) from Eq.~\ref{eqn:depinning:sw_Voltage}.  Due to the strong dependence on the random disorder-configuration the linear dependence is only realised on average over $20$ disorder configurations. The error-bars give the standard-deviation of the critical driving force in the  sample of disorder-configurations. }
\end{figure}

\tikzexternalenable
\begin{figure}[!tb]

\centering
\includegraphics{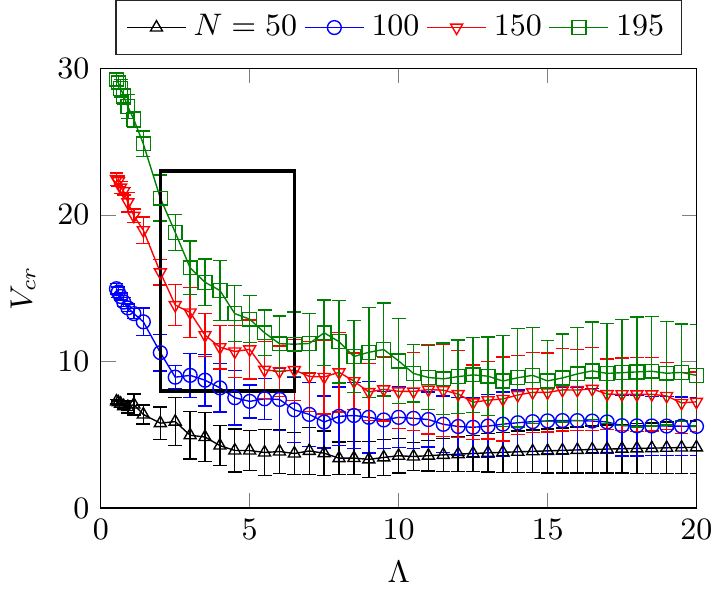}
\caption{\label{fig:depinning:Vt_lambda_disorder}(Color online) The critical driving force is plotted as a function of $\Lambda$ in disordered arrays for a wide range of $\Lambda$. For $\Lambda < 2$ the depinning-theory for the continuum limit is not applicable. For large $\Lambda$ the Larkin length $L_{p}$ is comparable to the chain length and $V_{cr}$ is independent of $\Lambda$, see also Eq.~\ref{eqn:depinning:sw_Voltag_short_array}. For intermediate $\Lambda$ the behaviour (black rectangle) is shown in Fig.~\ref{fig:depinning:Vt_lambda_disorder_small}  }
\end{figure}

\tikzexternalenable
\begin{figure}[!tb]
	\centering
\includegraphics{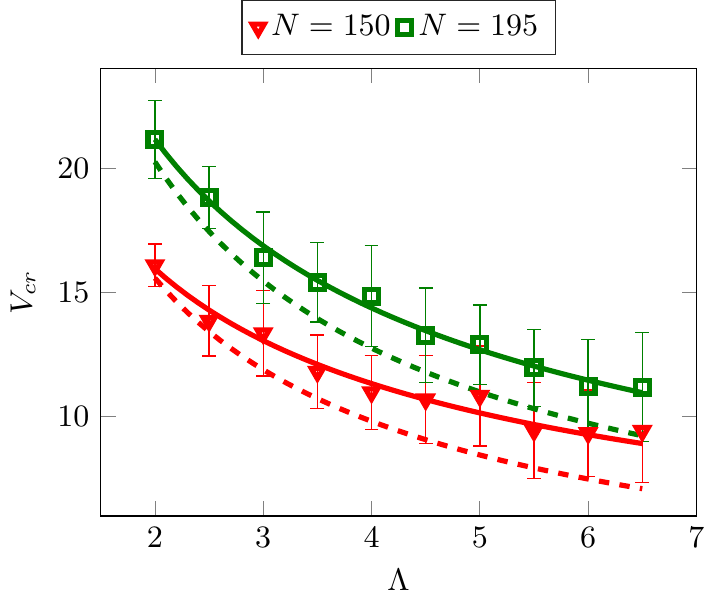}
	\caption{\label{fig:depinning:Vt_lambda_disorder_small} (Color online) A comparison of $V_{cr}$ in the intermediate $\Lambda$ regime of Fig.~\ref{fig:depinning:Vt_lambda_disorder} (black rectangle) with a fitted power-law decay (solid lines) and the analytic estimate Eq.~\ref{eqn:depinning:sw_Voltage} (dashed lines)  . From the fit  we obtain an exponent of $-0.49$ ($N=150$) and $-0.56$ ($N=195$)  while depinning theory predicts an exponent of $-\frac{2}{3}$.}
\end{figure}

\tikzexternalenable
\begin{figure}[!tb]
	\centering
\includegraphics{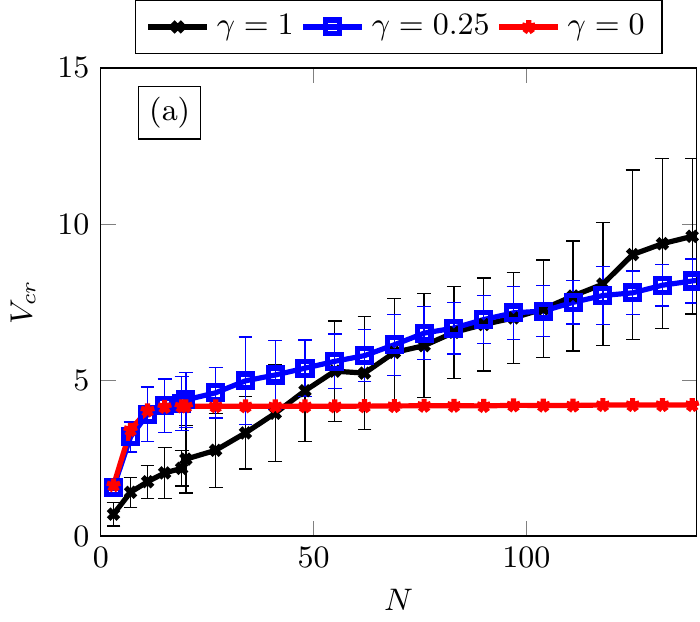}
\includegraphics{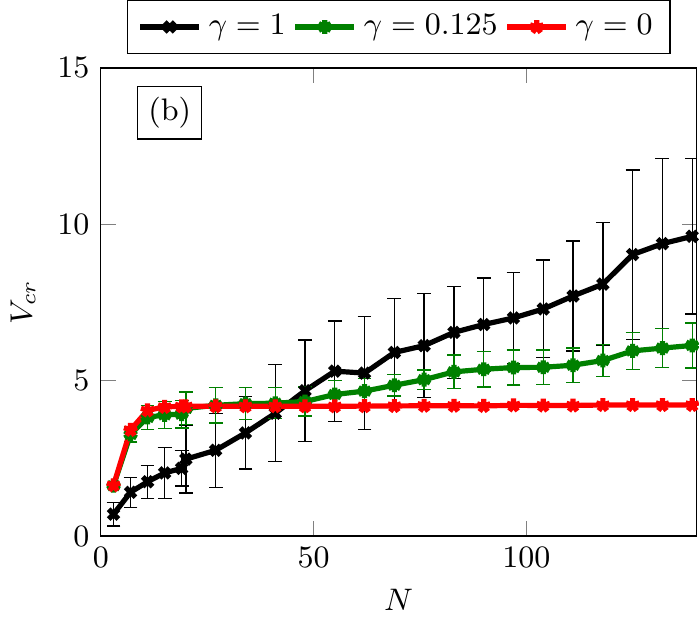}
	\caption{\label{fig:depinning:Vt_N_variable_disorder}(Color online) The critical driving force $V_{cr}$ is plotted as a function of the chain-length in a weak box-disorder model. To enhance visibility we show two subplots for different disorder strengths:  $\gamma = 0.25$, blue markers, subplot (a) and $\gamma = 0.125$ green markers subplot (b). For comparison the $V_{cr}$ of the clean case ($\gamma =0 $, red asterisk) and the maximally disordered model ($\gamma=1$, black crosses) are included in the plots. The behaviour of $V_{cr}$ changes when the chain length is equal to the correlation length $N=L_{corr}$. Below $N=L_{corr}$, the critical driving force has approximately the same value as in the clean case. Above $N=L_{corr}$ it increases linearly with $N$ as in the maximal disorder model, $L_{corr}(\gamma = 0.25) \approx 10$ and $L_{corr}(\gamma = 0.125) \approx 40$. }
\end{figure}

Here we present the critical driving force obtained from numerical simulations of the maximally disordered model. 
In Fig.~\ref{fig:depinning:Vt_N_disorder} we compare the dependence of  $V_{cr}$ on the parameter $N$  with analytic estimate 
in Eq.~\ref{eqn:depinning:sw_Voltage}. At large $N$, where the array is longer than the Larkin length $N > L_{p}$,  we find that the numerical simulations fit to the expected linear dependence on the system length. For small system lengths the switching voltage does not increase linearly with $N$, as expected in the saturation regime where the Larkin length is comparable to the system size (Eq.~\ref{eqn:depinning:sw_Voltag_short_array}).

The numerically determined dependence of $V_{cr}$  on $\Lambda$ is shown in Fig.~\ref{fig:depinning:Vt_lambda_disorder} and Fig.~\ref{fig:depinning:Vt_lambda_disorder_small}.  For small $\Lambda$ the inter-site distance is comparable to $\Lambda$ and the continuum limit of the standard depinning-picture does not apply. 
For large $\Lambda$ the Larkin-length is comparable to the chain-length $N$ and we observe a saturation of $V_{cr}$ with $\Lambda$. The saturation sets in for, 
\begin{align}
	N \approx \alpha_{sat} L_p \ ,
\end{align}
where $\alpha_{sat}$ is of order of one.  Comparing the analytic estimate Eq.~\ref{eqn:depinning:sw_Voltag_short_array} with the saturation points we expect $\alpha_{sat}$ in the range  $2.5 \le \alpha_{sat}  \le 3.5$. 

In this intermediate regime we expect a power-law behaviour with an exponent of $-\frac{2}{3}$ (Eq.~\ref{eqn:depinning:sw_Voltage}). Fitting the numerical data to a power-law we obtain the exponents $-0.49 \pm 0.05 $ ($N=150$) and $-0.56 \pm 0.03$ ($N=195$).  However this is limited by the numerically accessible  chain-lengths and we can not obtain a robust confirmation of the value of the exponent of $\Lambda$ from the numerical simulations.

\subsection{Weak disorder and emergent correlation length}
\label{sec:depinning:numeric:correlation}

To validate our analytic model of the introduction of a new length-scale by weak disorder, we also simulate the depinning-transition of the weakly disordered chain. We choose the disorder strengths $\gamma=0.25$, $L_{corr}(\gamma = 0.25) \approx 10$ and $\gamma = 0.125$, $L_{corr}(\gamma = 0.125) \approx 40$. In Fig.~\ref{fig:depinning:Vt_N_variable_disorder} it is shown that the system undergoes a transition when the array-length is equal to the correlation length, $N=L_{corr} $. Below $N < L_{corr} $ the chain is described by the clean chain model ($\gamma=0$). Above the transition the critical driving force increases linearly with $N$. The $N$-dependence of $V_{cr}$ matches the maximally disordered model $\gamma=1$. When the correlation length is significantly larger than the array size we can approximate all correlated disorder terms $F_i$ by a single value $F_i \approx F$. The perfectly correlated disorder term $F$ can be absorbed into the definition of the quasi-charge and the system is equivalent to the clean array without disorder $F_i = 0$.

When the length of the chain exceeds the correlation length one has to distinguish between two cases. The case when the correlation length is smaller than $\Lambda$ and the case where it is larger. The first case requires a careful treatment to map the weakly disordered case to an effective strongly disordered model. Here we limit ourselves to the simpler second case. In this case one can understand the behaviour of the critical driving force with the following simple argument. 

The typical length of a soliton $\Lambda$ is smaller than the correlation length and static solitonic solutions of the field $Q$ can exist in the chain. On the one hand this leads to the creation of a boundary soliton at the edge of the driven system that corresponds to the boundary soliton in the clean case. This gives rise to an offset critical driving force $V_{cr}^{\textrm{offset}}$. Since the chain is longer than $L_{corr}$ it can be subdivided into domains of length $L_{corr}$. To switch into the conduction regime, the applied driving force needs to overcome the transport threshold in each domain, where the transport threshold is proportional to the critical driving force in the clean chain Eq.~\ref{eqn:depinning:HDAnalyticEst},
\begin{align}
	V_{cr} -V_{cr}^{\textrm{offset}} &\propto  \sqrt{\frac{C}{2 C_{0}} V_Q^{max}} \frac{N}{L_{corr}} \label{eqn:depinning:WeakDisEst} \ .
\end{align}
This mechanism explains the linear increase in $V_{cr}$ 
seen in Fig.~\ref{fig:depinning:Vt_N_variable_disorder} for $N > L_{corr}$.

\section{Conclusions}
In this paper we have studied the depinning behaviour of discrete bosonic chain models that can be described by an effective Hamiltonian in the adiabatic limit that is similar to the sine-Gordon model. 
The most experimentally relevant realization of this model are linear arrays of Josephson junctions, however another possible realization is a ladder configuration of superconducting wires with quantum phase slip elements separating neighbouring superconducting loops.

We used analytical considerations and numerical simulations to determine the critical driving force required to overcome the pinning of bosons in the chain. In the parameter regime that corresponds to experimentally studied arrays we reproduce the recently observed behaviour\cite{Vogt2014b}. Going to new parameter regimes, namely short chains and weakly disordered chains, we see a saturation regime in short chains where the Larkin length exceeds the system length and the critical driving force is independent of the decay length $\Lambda$ of the repulsive interaction. In the weak disorder regime we observe the emergence of a new correlation length-scale $L_{corr}$. Both effects show good agreement between the analytic results and the numerical simulations 

\begin{acknowledgments}
We thank A.D. Mirlin, S.V. Syzranov, and T. Giamarchi for helpful discussions of the subject. This work was funded by the Russian Science Foundation under Grant No. 14-42-00044.
\end{acknowledgments}

\bibliography{Promotion-numeric_paper}

\end{document}